\documentstyle [12pt]{article}
\def\int {\intop \limits}

\def\fnote#1{\footnote}
\begin{document}
\newcommand{\dst}[1]{\displaystyle{#1}}
\newcommand{\barl}{\begin{array}{rl}}
\newcommand{\ball}{\begin{array}{ll}}
\newcommand{\ear}{\end{array}}
\newcommand{\barc}{\begin{array}{c}}
\newcommand{\sne}[1]{\displaystyle{\sum _{#1} }}
\newcommand{\sn}[1]{\displaystyle{\sum ^{\infty }_{#1} }}
\newcommand{\ini}[1]{\displaystyle{\int ^{\infty }_{#1}}}
\newcommand{\myi}[2]{\displaystyle{\int ^{#1}_{#2}}}
\newcommand{\inn}{\displaystyle{\int }}
\newcommand{\be}{\begin{equation}}
\newcommand{\ee}{\end{equation}}
\newcommand{\aq}[1]{\label{#1}}

\vspace*{4.0cm}
\centerline{\Large {\bf On Basic Concepts of the Quasiclassical}}
\vskip .25cm
\centerline{\Large {\bf Operator Approach}}
\vskip .5cm
\centerline{\large{\bf V. N. Baier and V. M. Katkov}}
\centerline{Budker Institute of Nuclear Physics,
 630090 Novosibirsk, Russia}
\vskip 2.0cm
\begin{abstract}
We present a derivation of the probability of bremsstrahlung
from high-energy electrons in a screened Coulomb field using
the quasiclassical operator method.
It is shown that recent Zakharov's criticism is completely
groundless and comes from misunderstanding of the method.
We confirm all our published results.
\end{abstract}

\newpage
1. Comment, recently published by Zakharov \cite{1}, shows that
it is reasonable to discuss once more the basis of the quasiclassical
operator (QO) method developed by authors in
\cite{2}, \cite{3}. The method was presented in detail in the book \cite{4}
and later in the books \cite{5} and \cite{6} (the latter is the enlarged
English translation of the book \cite{5}). Very clear description of the
QO method is given in the popular textbook \cite{7}.

This paper is divided into two parts. In the first part we 
discuss in frame of the QO method the derivation 
of the basic formulae describing the radiation process
paying special attention to the important details
of the approach. In the second part we analyze Zakharov's claims step
by step and show that all his pretensions are erroneous because of
complete misunderstanding of the QO method.

In this paper we consider the applicability of the QO method
to the problem of radiation from a ultrarelativistic particle at a potential
scattering. In this process the emitted photon and the final electron are
moving at a small angle to the initial electron momentum and the large
angular momenta $l \gg 1$ contribute. In this situation the
quasiclassical scattering theory is applicable. 
The very important result of the method is that recoil at radiation
is incorporated into the theory in the {\em universal} form for
any external field. There are two essentially different cases in
application of the QO method. In the first case scattering in a
potential can be considered in {\em classical} terms:
phase shifts are large, there is a correspondence between
the impact parameter and the momentum transfer. Therefore it
is possible to use the
version of the QO method where one can instead of operators
substitute classical variables in the {\em coordinate} space
in an expression for probability of the process
(see, \cite{2}, Secs.9-15
in \cite{4}, Secs.2-5 in \cite{5}, \cite{6}, Sec.90 in \cite{7}).
In the second case, the process of scattering can't be described in
classical terms. However, for scattering of ultrarelativistic
particles where large angular momentum contributes, one can use
the {\em quasiclassical} approximation for description of scattering
including the situation where the phase shifts are small
(see \cite{3}, Secs.9,16-20
in \cite{4}, Secs.2,7 in \cite{5}, \cite{6}, Sec.96 in \cite{7}).
Precisely this formulation of the method 
must be applied for consideration of radiation from
ultrarelativistic particles at scattering on atoms in a media.
In our recent papers devoted to the theory of Landau-Pomeranchuk-Migdal
(LPM) effect \cite{7a}-\cite{7d} we use the mentioned formulation of 
the QO method.
Therefore we will analyze it in detail below.

2.Because the LPM effect can be observed at very high energies, we can
consider the case of complete screening, so that (we employ units
such that $\hbar=c=1$)
\begin{equation}
a_s \ll \frac{1}{q_{min}}=\frac{2\varepsilon(\varepsilon-\omega)}{\omega m^2}
\equiv l_0,
\label{1}
\end{equation}
where $a_s$ is the screening radius $(a_s \simeq 111Z^{-1/3}\lambda_c,
\lambda_c =1/m), Z$ is the charge of a nucleus, $q_{min}$ is the
minimal momentum transfer which is longitudinal (with respect to the momentum
of the initial electron ${\bf p}$), $\varepsilon (\omega)$
is the energy of the initial electron (the emitted photon), 
$m$ is the electron mass,
$l_0$ is the radiation formation length for a small angle
scattering on an isolated atom.
Note that in frame of the QO method the radiation problem is
solved for the case of an arbitrary screening , see Sec.18
in \cite{4}. The impact parameters $\varrho$,
contributing into the scattering cross section, are small comparing the
formation length ($\varrho \leq a_s \ll l_0$) in a screened Coulomb potential.
This means that the scattering of the ultrarelativistic particles
(the virtual electron is close to the mass shell) takes place independently
of radiation process (see, Sec.7 in \cite{5}, \cite{6}, Sec.96 in \cite{7}).
Thus, we can present the cross section of radiation as a product of the
probability of emission of a photon with the momentum ${\bf k}$ at given
momentum transfer ${\bf q_{\perp}}~({\bf q_{\perp}p}=0)$,
and the cross section of scattering
$d\sigma({\bf q_{\perp}})$
of a particle with the same momentum transfer ${\bf q_{\perp}}$:
\begin{equation}
d\sigma_{\gamma}=W_{\gamma}({\bf q_{\perp}}, {\bf k})d^3k
d\sigma({\bf q_{\perp}}).
\label{2}
\end{equation}
We show below that in frame of the QO method the probability of radiation
$W_{\gamma}({\bf q_{\perp}}, {\bf k})$ is given by the {\em classical}
trajectory of a particle in "the form of an angle"
in the {\em momentum space}
\begin{equation}
{\bf p}(t)=\vartheta(-t){\bf p}+\vartheta(t)({\bf p}+{\bf q_{\perp}}),
\label{3}
\end{equation}
while the cross section $d\sigma({\bf q_{\perp}})$ should be taken
in the {\em eikonal} form.

3. The probability of photon emission from an electron in frame of
the QO method has the form (see, Eqs.(16.7)-(16.12) in \cite{4},
Eqs.(7.1)-(7.12) in \cite{5}, \cite{6}, Eqs.(96.1)-(96.8) in \cite{7})
\begin{equation}
dw=<i|M^{+} M|i>,
\label{4}
\end{equation}
where
\begin{equation}
M=\frac{e}{2\pi \sqrt{\omega}}\int_{-\infty}^{\infty}R(t)
\exp \left[i\int_{0}^{t} \frac{kp(t')}{\varepsilon-\omega} dt' \right],
\label{5}
\end{equation}
here $R(t)=R({\bf p}(t))$, $R({\bf p})$ is the
matrix element for the free particles depending on the electron spin,
$kp=\omega {\cal H}-{\bf k}{\bf p}, {\cal H}=\sqrt{{\bf p}^2+m^2}$,
$|i>$ is the state vector of the initial particle at the time $t=0$,
and ${\bf p}(t)$ is the operator of momentum in the Heisenberg picture:
\begin{equation}
{\bf p}(t)=\exp (-iHt) {\bf p} \exp (iHt),\quad
H={\cal H}+V({\bf r}),
\label{6}
\end{equation}
where $V({\bf r})$ is the potential of an atom.

We present the evolution operator as
\begin{equation}
\exp (-iHt) =\exp (-i{\cal H}t) N(t),\quad
N(t)=\exp (i{\cal H}t) \exp (-i({\cal H}+V)t).
\label{7}
\end{equation}
Differentiating the last expression over the time we obtain
\begin{equation}
\frac{dN(t)}{dt} =-i\exp (i{\cal H}t) V({\bf r}) \exp (-i({\cal H}+V)t)=
-iV({\bf r}+{\bf v}t)N(t), {\bf v}=\frac{{\bf p}}{{\cal H}}.
\label{8}
\end{equation}
The solution of this differential equation for the initial condition
$N(0)=1$ is
\begin{equation}
N(t)={\rm T} \exp \left[-i\int_{0}^{t} V({\bf r}+{\bf v}t') dt'\right],
\label{9}
\end{equation}
where T is the operator of the chronological product. Bearing in mind that
the commutator
\begin{equation}
\left[ r_i, v_j \right] = \frac{i}{{\cal H}}(\delta_{ij}-v_iv_j),
\label{10}
\end{equation}
one can drop operator T of the chronological product
within relativistic accuracy (i.e. with accuracy up to terms
$\sim 1/\gamma$) and present operator $N(t)$ as
\begin{equation}
N(t) \simeq \exp
\left[ -i\int_{0}^{t} V(\mbox{\boldmath$\varrho$}, z+t')dt' \right],
\label{11}
\end{equation}
where the axis $z$ is directed along the momentum of the initial particle.
If the formation time of radiation is much longer
than the characteristic time of the scattering, one can present
the dependence of the operator ${\bf p}(t)$ on the time in Eq.(\ref{5}) as
\begin{equation}
{\bf p}(t) =\vartheta(-t) {\bf p}(-\infty)+\vartheta(t) {\bf p}(\infty),
\quad {\bf p}(\pm \infty) = N^+(\pm \infty){\bf p}N(\pm \infty).
\label{12}
\end{equation}
It should be pointed out that in the case when the scattering process
is of nonclassical character the operators ${\bf p}(-\infty)$ and
${\bf p}(\infty)$ are noncommutative among themselves and,
generally speaking, one can't neglect their commutator. Let us mention also
that using Eq.(\ref{11}) one find
\begin{equation}
{\bf p}_{\perp}(-\infty) \simeq {\bf p}_{\perp}+\int_{-\infty}^{z}
\mbox{\boldmath$\nabla$}_{\varrho} V(\mbox{\boldmath$\varrho$}, z')dz',\quad
{\bf p}_{\perp}(\infty) \simeq {\bf p}_{\perp}-\int_{z}^{\infty}
\mbox{\boldmath$\nabla$}_{\varrho} V(\mbox{\boldmath$\varrho$}, z')dz'.
\label{13}
\end{equation}
These approximate expressions are written in the classical form
when the entering operators commutate among themselves. Substituting the
"trajectory" (\ref{12}) in Eq.(\ref{5}) we obtain (this is Eq.(7.7) in
\cite{5}, \cite{6})
\begin{equation}
M=\frac{ie(\varepsilon-\omega)}{2\pi \sqrt{\omega}}
\left[\frac{R({\bf p}(\infty))}{kp(\infty)}-
\frac{R({\bf p}(-\infty))}{kp(-\infty)} \right].
\label{14}
\end{equation}

The state $|i>$ in Eq.(\ref{4}) denotes the wave function
in the configuration space in the time $t=0$. However,
in the scattering problem under consideration (motion of a particle in
a local potential), the wave function 
are defined at $t \rightarrow \pm \infty$. 
Let us consider the interconnection between these wave
functions. By definition of the state evolution, we have
\begin{eqnarray}
&& \lim_{t \rightarrow \infty} \exp (-i{\cal H}t) |\infty>=
\lim_{t \rightarrow \infty} \exp (-iHt) |0>, \nonumber \\
&&|0>=N^+(\infty)|\infty> \equiv |{\rm out}> \simeq \exp
\left[i\int_{z}^{\infty} V(\mbox{\boldmath$\varrho$}, z')dz'
\right] |\infty>.
\label{15}
\end{eqnarray}
For the state evolving from $t \rightarrow -\infty$ we have
\begin{equation}
|0>=N^+(-\infty)|-\infty> \equiv |{\rm in}> \simeq \exp
\left[-i\int_{-\infty}^{z} V(\mbox{\boldmath$\varrho$}, z')dz'
\right] |\infty>.
\label{16}
\end{equation}
As a rule, the state at $t \rightarrow \pm \infty$ are the plane waves,
i.e. they are the eigenstates of the momentum operator {\bf p}.
However, one can use also another definition of these states, e.g.
wave packets. Such definition was used in description of the bremsstrahlung
at the collision of two high-energy beams with the restricted transverse
dimensions \cite{6a}. If there is a summation over the set of states, i.e.
they are the intermediate states,
choice of the states is determined by a convenience of calculation.
This circumstance was used in \cite{9} for the derivation of connection
between spectra of the bremsstrahlung and the pair creation in the 
theory exact in $Z\alpha$.

4.It follows from the above analysis that for the derivation of
the differential cross section of the bremsstrahlung
in Eqs.(\ref{2}), (\ref{4}),
it is necessary to insert  the projection operator $|f><f|$, which corresponds "out"
states of the plane wave with the momentum ${\bf p}_f$,
between the operators
$M^+$ and $M$, and to take "in" states with the momentum ${\bf p}_i$
as the states $|i>$. So,
the initial state is the eigenvector of the operator ${\bf p}(-\infty)$
and the final state is the eigenvector of the operator ${\bf p}(\infty)$:
\begin{eqnarray}
&&{\bf p}(-\infty) |i> =N^+(-\infty){\bf p}N(-\infty)N^+(-\infty)|{\bf p}_i>
={\bf p}_i|i>, \nonumber \\
&&{\bf p}(\infty) |f> =N^+(\infty){\bf p}N(\infty)N^+(\infty)|{\bf p}_f>
={\bf p}_f|f>.
\label{17}
\end{eqnarray}
Using (\ref{17}) we deduce for the matrix element of the
operator M (\ref{14})
\begin{equation}
M=\frac{ie(\varepsilon-\omega)}{2\pi \sqrt{\omega}}
\left[\frac{R({\bf p}_f)}{kp_f}-
\frac{R({\bf p}_i)}{kp_i} \right] <f|i>.
\label{18}
\end{equation}
Using the definitions of "in" and "out" states Eqs.(\ref{15}),(\ref{16})
we have
\begin{eqnarray}
&& <f|i>=<{\bf p}_f|N(\infty)N^+(-\infty)|{\bf p}_i>=
<{\bf p}_f|{\rm T} \exp \left[-i\int_{-\infty}^{\infty} V({\bf r}
+{\bf v}t) dt\right]|{\bf p}_i> \nonumber \\
&& \simeq <{\bf p}_f| \exp \left[-i\int_{-\infty}^{\infty} V({\bf r}
+{\bf v}t) dt - \frac{1}{2}\int_{-\infty}^{\infty}dt_2
\int_{-\infty}^{\infty}dt_1 [V(t_2), V(t_1)] \right]|{\bf p}_i>,
\label{19}
\end{eqnarray}
where $V(t_1)$ means $V({\bf r}+{\bf v}t_1)$, etc.
We retain here only the first
two terms in the decomposition of the T-product being restricted to
the commutator
\begin{equation}
[V(t_2), V(t_1)] = -\frac{i}{{\cal H}}(t_2-t_1)
\mbox{\boldmath$\nabla$}_{\perp} V(t_2)
\mbox{\boldmath$\nabla$}_{\perp} V(t_1).
\label{20}
\end{equation}
The first term in the decomposition of the T-product in Eq.(\ref{19})
gives the scattering amplitude in the eikonal approximation,
the second term gives the correction to the eikonal approximation:
\begin{equation}
<f|i>=\int_{}^{}d^2\varrho \exp \left[i{\bf q}_{\perp}
\mbox{\boldmath$\varrho$} +i\chi (\mbox{\boldmath$\varrho$}) \right]
2\pi \delta(p_{f\parallel}-p_{i\parallel}),
\label{21}
\end{equation}
where
\begin{eqnarray}
&& \chi (\mbox{\boldmath$\varrho$})=\chi_0(\mbox{\boldmath$\varrho$})
+\chi_1(\mbox{\boldmath$\varrho$}),\quad \chi_0(\mbox{\boldmath$\varrho$})
=-\int_{-\infty}^{\infty} V(\mbox{\boldmath$\varrho$}, z)dz \nonumber \\
&& \chi_1(\mbox{\boldmath$\varrho$})=\frac{1}{2\varepsilon}
\int_{-\infty}^{\infty} dz_2 \int_{-\infty}^{\infty} dz_1
\mbox{\boldmath$\nabla$}_{\perp} V(z_2)
\mbox{\boldmath$\nabla$}_{\perp} V(z_1)(z_2-z_1)\vartheta(z_2-z_1).
\label{22}
\end{eqnarray}
In the centrally symmetric potential we have
\begin{equation}
\chi_0(\varrho)=-\int_{-\infty}^{\infty} V(\sqrt{\varrho^2+z^2})dz,\quad
\chi_1(\varrho)=-\frac{\varrho^2}{\varepsilon}\int_{-\infty}^{\infty}
\frac{\partial V^2(\sqrt{\varrho^2+z^2}) }{d\varrho^2}dz
\label{23}
\end{equation}
In the Coulomb field $\chi_1=(Z\alpha)^2\pi/(2\varrho \varepsilon)$.
The main contribution into corrections to the probability 
of the bremsstrahlung connected with the phase $\chi_1$ gives 
the impact parameters $\varrho \sim \lambda_c=1/m$,
so that $\chi_1 \sim (Z\alpha)^2/\gamma$. The corresponding corrections 
$\sim 1/\gamma$ to the
bremsstrahlung cross section were calculated by authors in \cite{6b}.

We introduce now the notations ${\bf p}_i \equiv {\bf p},~{\bf p}_f=
{\bf p}'+{\bf k}$ where ${\bf p}'$ is the momentum of electron after
photon emission. Then neglecting the terms of the order of
$q_{\parallel}$ in the argument of the $\delta$-function in Eq.(\ref{21})
we have in the region $q_{\perp} \gg q_{\parallel}$ which contributes for
the potential considered 
\begin{equation}
\delta(p_{f\parallel}-p_{i\parallel}) \simeq
\delta(\varepsilon'+\omega-\varepsilon), \quad \varepsilon'=
\sqrt{{\bf p}'^2+m^2}.
\label{24}
\end{equation}
Using this relation one can express the propagator $kp_f$ through the
propagator $kp'$ (up to terms of the order $1/\gamma^2$)
\begin{eqnarray}
&& kp_f=\omega\sqrt{({\bf p}'+{\bf k})^2+m^2}-{\bf k}({\bf p}'+{\bf k})
\nonumber \\
&&=\omega \sqrt{\varepsilon^2-2kp'}+kp'-\omega \varepsilon
\simeq \omega \varepsilon \left(1-\frac{kp'}{\varepsilon^2} \right) +
kp'-\omega \varepsilon = \frac{\varepsilon'}{\varepsilon}kp'.
\label{25}
\end{eqnarray}
Our final result for the differential probability of radiation in
Eq.(\ref{2}) is therefore
\begin{equation}
W_{\gamma}({\bf q_{\perp}}, {\bf k}) = \frac{\alpha}{(2\pi)^2}
\frac{1}{\omega}
\left|\frac{\varepsilon R({\bf p}'+{\bf k})}{kp'}-
\frac{\varepsilon' R({\bf p})}{kp} \right|^2
\label{26}
\end{equation}
The explicit expression for the probability $W_{\gamma}({\bf q_{\perp}}, 
{\bf k})$ with regard for polarization and spin effects 
are given in \cite{4}.

5. In the above analysis we traced in detail the transition  
to the expressions
calculated on the trajectory of particle in "the form of an angle"
in the momentum space. Precisely these {\em definite} trajectories (taking
into account the recoil at the radiation) determine the probability of
photon emission
\begin{eqnarray}
&& dw=\frac{\alpha}{(2\pi)^2} \frac{d^3k}{\omega}
\int_{}^{}dt_2\int_{}^{}dt_1 R^{\ast}(t_2) R(t_1)
\exp\left[-\frac{i\varepsilon}{\varepsilon'}\int_{t_1}^{t_2}
kv(t) dt \right]
\nonumber \\
&&= W_{\gamma}d^3k,\quad v=\frac{p}{\varepsilon} = (1, {\bf v}),\quad
\varepsilon'=\varepsilon-\omega.
\label{27}
\end{eqnarray}

When the projectile is moving in a medium it scatters on atoms.
In this case the probability should be averaged over all possible
trajectories with the weight function defined by the cross section
$d\sigma({\bf q}_{\perp})$. In the presence of a macroscopic external field
there is the systematic variation of the mean velocity of a projectile.
The probability of radiation in this case is given by the same Eq.(\ref{27}).
 Both, the systematic variation
of the mean velocity of the projectile as well as the fluctuations of
it velocity due to scattering with respect to the mean value
are taken into account 
with the aid of the distribution function.
Thus, when the projectile is moving in a medium the probability
of photon emission per unit time has the form (see, Eq.(20.31) in
\cite{4} and Eq.(2.4) in \cite{4a}) 
\begin{eqnarray}
&& dw=\left<\frac{dw}{dt} \right>=\frac{\alpha}{(2\pi)^2} \frac{d^3k}{\omega}
{\rm Re}\int_{0}^{\infty}d\tau \exp\left(-i\frac{\varepsilon}{\varepsilon'}
\omega \tau \right)\int_{}^{}d^3v \int_{}^{}d^3v'
\nonumber \\
&& \times  \int_{}^{}d^3r \int_{}^{}d^3r'
{\cal L}(\mbox{\boldmath$\vartheta$}, \mbox{\boldmath$\vartheta$}')
F_i({\bf r}, {\bf v}, t) F_f({\bf r}', {\bf v}', \tau; {\bf r}, {\bf v})
\exp\left[i\frac{\varepsilon}{\varepsilon'}
{\bf k}({\bf r}'-{\bf r}) \right].
\label{28}
\end{eqnarray}
We consider here the case of an infinitely thick target.
The distribution functions $F_i$ and $F_f$ satisfy
the kinetic equation (see, Eq.(20.32) in
\cite{4} and Eq.(2.5) in \cite{4a})):
\begin{eqnarray}
&& \frac{\partial F({\bf r}, {\bf v}, t)}{\partial t}+
{\bf v}\frac{\partial F({\bf r}, {\bf v}, t)}{\partial {\bf r}}+
{\bf w}\frac{\partial F({\bf r}, {\bf v}, t)}{\partial {\bf v}}
\nonumber \\
&& = n\int_{}^{}\sigma({\bf v}, {\bf v'})
\left[F({\bf r}, {\bf v}', t)-F({\bf r}, {\bf v}, t) \right]d^3v',
\label{28a}
\end{eqnarray}
where $n$ is the number density of atoms in the medium,
and $\sigma({\bf v}, {\bf v'})$ is the scattering cross section
which in our case should be
calculated in the eikonal approximation in the screened Coulomb potential.
The combinations entering Eq.(\ref{27}) are
\begin{eqnarray}
&& \int_{t_1}^{t_2} kv(t) dt = \int_{t_1}^{t_2} (\omega-{\bf kv}(t)) dt
= \omega \tau -{\bf k}({\bf r}(t_2)-{\bf r}(t_1))
\rightarrow \omega \tau -{\bf k}({\bf r}'-{\bf r}),
\nonumber \\
&& R^{\ast}(t_2) R(t_1) = \frac{1}{2\varepsilon'^2}
\left[\frac{\omega^2m^2}{\varepsilon^2}+(\varepsilon^2+\varepsilon'^2)
{\bf v}_{\perp}(t_2){\bf v}_{\perp}(t_1) \right]
\nonumber \\
&& \rightarrow \frac{1}{2\varepsilon'^2}
\left[\frac{\omega^2m^2}{\varepsilon^2}+(\varepsilon^2+\varepsilon'^2)
\mbox{\boldmath$\vartheta$} \mbox{\boldmath$\vartheta$}' \right]
\equiv \frac{1}{2}
{\cal L}(\mbox{\boldmath$\vartheta$}, \mbox{\boldmath$\vartheta$}').
\label{29}
\end{eqnarray}

Just this approach was used in papers \cite{7a}-\cite{7d}
in the derivation of the principal equations in the theory of the LPM
effect.
\vskip10mm

Let us analyze specific Zakharov's claims in p.4 of \cite{1} (given
in quotation marks).
\begin{enumerate}
\item ".. Eq.(1) (this is our Eq.(\ref{27})) derived
neglecting the variation of the field
acting on electron...". This statement is wrong. The Eq.(\ref{27}),
is valid
for any variation of the external field. Although we consider
the case of slowly varying external field in the first derivation \cite{2},
nevertheless for $\gamma \gg 1$ (this is the only case we consider),
this result can used for the Coulomb field
(see Eqs.(10) and (11) in \cite{3}). Later we show that Eq.(\ref{27}),
is valid for any field (see \cite{5}, \cite{6}) including the field of
the plane wave where the dependence on both coordinate and time is
essential.
\item "..In the quantum regime the above correspondence
between the scattering angle and the electron impact parameter
in interaction with a medium consistent is lost. In this case
one must take into account accurately the variation of the field acting
on the electron in evaluating the radiation rate. \underline {Eq.(1)
(obtained
for a smooth field) does not make any sense in the}
\underline {quantum regime.}
It is well known that for a screened Coulomb potential
the quasiclassical situation takes place at $Z\alpha \gg 1$. In real media,
when $Z\alpha < 1$ we have essentially quantum regime. For this reason,
it is evident that Eqs.(1) and (2)  (Eq.(2) we will discuss below) are not
justified for real media. The use of Eqs.(1) and (2) which are not
valid in the quantum regime is the main conceptual error in the treatment
of the LPM effect given in Refs.[11, 12]. (here this is Refs.\cite{4} and
\cite{4a})."

A few errors  accumulate here. First of all it is evident that Zakharov
is not familiar with classical papers of Bethe and Maximon \cite{8} (see also
\cite{9}-\cite{11})
where cross sections of the bremsstrahlung and the pair creation in the
Coulomb field for high energy region are calculated with
the use of the quasiclassical
wave functions. The results obtained in the mentioned papers are valid
for any value of $Z\alpha$, for low $Z$ they coincide with the Bethe-Heitler
cross section, for $Z\alpha \sim 1$ the cross sections contain additional
term which is the Coulomb correction. We want to stress that our
approach has the same region of applicability as the Bethe-Maximon theory.
This theory is developed in frame of Schr\"odinger picture, while we are
developed approach in frame of Heisenberg (operator)
picture. Because of this,
starting with some stage of calculation our results coincide with those
of Bethe and Maximon or Olsen, Maximon, and Wergeland \cite{10}. The case
of a screened Coulomb field is considered in the last paper.

Second, as it is shown above, our approach is basically
quantum (quasiclassical) and only the final result can be in many cases,
expressed in classical terms as it was explained above (and in
\cite{2}-\cite{7}). Moreover, in the papers \cite{3}, \cite{7}
the case of the Coulomb field was considered. 
For calculation of the probability of radiation
in this field for $q \sim q_{min}$ one can't use 
the trajectory "in the form of an angle", the impact parameters
$\varrho \sim l_0$ contribute (see Eq.(\ref{1})), and the cross section
of radiation can't be presented in the form Eq.(\ref{2}). But even 
in this case the Eqs.(\ref{4}) and (\ref{5}) are valid.
The underlined
claim is erroneous. In fact, the quantum-mechanical properties of scattering
are relevant part of our theory.
\item point 2 "The authors of Refs.[11,12] (this is Refs.\cite{4}, \cite{4a}
here) compensate the incorrectness of Eq.(1) (this is Eq.(\ref{27}) here) by
another error in the procedure of averaging over the electron trajectories.
According to their logic the QO expression (1) (this is Eq.(\ref{27}) here)
must be averaged over all possible classical trajectories (and author say
this)." The last statement has nothing common with our method. 
As it was shown above the Eq.(\ref{27}) is valid in all cases when the motion
of particle can be described by quasiclassical quantum mechanics.
As was explained above, the averaging procedure appears only
in the case when we have to consider the
scattering of the projectile in a medium. In this situation
the radiation process is described by Eq.(\ref{28}). In a very definite
form this statement is contained in \cite{4a} (the paragraph before Eq.(2.4))
\end{enumerate}

Let us make the final outlook on the discussion.
\begin{itemize}
\item The method developed in \cite{2}-\cite{3} and presented in detail
in \cite{4}-\cite{7} is basically quantum theory in Heisenberg
picture in quasiclassical approximation. So, in use of the theory one have to
proceed with operator calculus. In many case, but not always,
the final result can be expressed in classical terms (also in a very definite
form).  Of course, the possibility to use classical form is a great advantage
of the method since it simplifies essentially a solution of a particular
problem and permits to solve problems which could not be solved by other
methods.
\item Zakharov \cite{1} expressed pretensions to Eq.(2.1) of \cite{4a}.
This formula is given as known with the reference to Eq.(9.27) of
\cite{4}. So, to use this equation one have to read, at least, the
mentioned Sec.9 of this book. The problem was considered in detail also
in books \cite{5}, \cite{6} which contain derivation of the mentioned
formula. However these books were not mentioned
in \cite{1} as well as known textbook \cite{7}. 
Instead of this Zakharov invents his own prescriptions
which have nothing common with our method and then criticizes these
prescriptions. See, e.g. p.4, point 2 "...the distribution function...
should satisfy the kinetic equation in which the collisional term contains
the {\em classical} scattering cross section".
\item It is evident that there is essential difference between the
classical theory and the quantum theory in quasiclassical approximation.
Unfortunately, from Comment \cite{1} it looks that Zakharov don't distinguish
between these two.
\item It is curious that any references
on incorrect expressions in our publications are absent in Comment \cite{1}.
\item It is rather strange to read that  in 1996
Zakharov in \cite{12} quite successfully derived formulae,
in absence of an external field, contained in our
paper \cite{4a} and book \cite{4}. Moreover, he explains what kind
of cross section should be substituted (just the formulae used in
 the mentioned publications).
\item It should be noted that although Zakharov 
reproduced some of our results many years later than they were obtained
in original papers,
he never refer to our publications before
and only in Comment \cite{1} there is citing of our publications.
\item It's amusing that in list of criticized papers there is paper
\cite{13} devoted to quantum theory of the transition radiation
where there is no scattering at all (the criticism devoted
to description of scattering) and this once more shows the level of the
pretensions.
\end{itemize}

\end{document}